\documentclass{amsart}

\usepackage{amsmath,amsthm,amsfonts,amssymb,amscd,amsbsy}
\usepackage{graphicx}
\usepackage{array}
\usepackage{enumerate}
\usepackage[nodayofweek]{datetime}
\usepackage[english]{babel}
\usepackage{mathtools}
\usepackage[toc,page]{appendix}
\usepackage{verbatim} 
\usepackage{amsthm} 
\usepackage{enumitem}
\usepackage{enumerate}
\usepackage{bbm} 
\usepackage[normalem]{ulem} 
\usepackage{bm}
\usepackage{hyperref}

\newcommand{\be}{\begin}
\newcommand{\e}{\end}
\newcommand{\beq}{\begin{equation}}
\newcommand{\eeq}{\end{equation}}

\renewcommand{\l}{\left}
\renewcommand{\r}{\right}
\renewcommand{\d}{\mathrm{d}} 

\newcommand{\set}[1]{\mathbb{#1}}
\newcommand{\curly}[1]{\mathcal{#1}}

\newcommand{\setof}[2]{\left\{ #1\; : \;#2 \right\}}

\newcommand{\R}{\set{R}}
\newcommand{\C}{\set{C}}
\newcommand{\Z}{\set{Z}}

\newcommand{\T}{\set{T}}
\newcommand{\om}{\omega}

\newcommand{\eps}{\epsilon}

\newcommand{\lam}{\lambda}

\newcommand{\Gam}{\Gamma}

\newcommand{\al}{\alpha}

\newcommand{\ttmatrix}[4]{\left(\be{array}{cc} #1&#2\\	#3&#4 \e{array}	\right)}
\newcommand{\tvector}[2]{\left(\be{array}{c}#1\\#2\e{array}\right)}

\theoremstyle{definition}

\theoremstyle{remark}

\begin{document}

\title[Finite-size Lyapunov exponent for the skew-shift potential]{On the finite-size Lyapunov exponent for the Schr{\"o}dinger operator with skew-shift potential}
\date{April 18, 2019}

\author{Paul Michael Kielstra}
\email{pmkielstra@college.harvard.edu}
\author{Marius Lemm}
\email{mlemm@math.harvard.edu}
\address{Department of Mathematics, Harvard University, 1 Oxford Street, Cambridge, MA 02138, USA}

\maketitle

\begin{abstract}
It is known that a one-dimensional quantum particle is localized when subjected to an arbitrarily weak random potential. It is conjectured that localization also occurs for an arbitrarily weak potential generated from the nonlinear skew-shift dynamics: $v_n=2\cos\l(\binom{n}{2}\omega +ny+x\r)$ with $\om$ an irrational number. Recently, Han, Schlag, and the second author derived a finite-size criterion in the case when $\omega$  is the golden mean, which allows to derive the positivity of the infinite-volume Lyapunov exponent from three conditions imposed at a fixed, finite scale. Here we numerically verify the two conditions among these that are amenable to computer calculations. \end{abstract}

\section{Introduction}

A one-dimensional quantum particle living on $\Z$ with energy $E\in \R$ is described by the discrete Schr{\"o}dinger equation
\beq\label{eq:schr}
\psi_{n+1}+\psi_{n-1}+\lam v_n\psi_n=E\psi_n,
\eeq
where $\psi=(\psi_n)_{n\in\Z}$ is a sequence in $\ell^2(\Z;\C)$. The real-valued potential sequence $v=(v_n)_{n\in \Z}$ represents the environment that the particle is subjected to. (The ``coupling constant'' $\lam>0$ is factored out for convenience.) As first famously realized by Anderson in 1958 \cite{Anderson}, the decoherence introduced by a random (meaning independent and identically distributed) sequence of potentials can drastically affect the spectral and dynamical properties of the quantum particle. Physically, one observes a sudden onset of insulating behavior in the presence of a random environment (``Anderson localization''). Mathematically, it is known that for arbitrarily small $\lam>0$, the one-dimensional Schr{\"o}dinger operator 
\beq\label{eq:Hdefn}
(H\psi)_n=\psi_{n+1}+\psi_{n-1}+\lam v_n\psi_n.
\eeq
has pure point spectrum with exponentially decaying eigenfunctions  \cite{CKM,GMP,KS}.

A natural follow-up question is then: How random does the environment have to localize the quantum particle? Alternative ``quasi-random'' environments are generated by sampling a nice function along the orbit of an ergodic dynamical system. This question is interesting from a purely mathematical ergodic theory perspective, but it also has practical implications, since computer simulations are mostly based on appropriate pseudo-random number sequences. For example, one can consider $v_n=2\cos(n\al+\theta)$ generated from sampling cosine along an irrational circle rotation; this is the well-known Harper (or Almost-Mathieu) model . It turns out that these linear underlying dynamics only produce localization for sufficiently strong potentials, namely only for $\lam>1$ \cite{Jit}. One would thus like to consider dynamics which are slightly more quasi-random than the shift.

A standing conjecture in this direction concerns the case when the potential is generated from the nonlinear skew-shift dynamics $T:\mathbb T^2\to\mathbb T^2$, $T(x,y)=(x+y,y+\omega)$, namely it is of the form
\beq
\label{eq:vndefn}
v_n=2\cos\l(\binom{n}{2}\omega +ny+x\r)
\eeq
with $\omega$ irrational (say Diophantine). The key difference between \eqref{eq:vndefn} compared to $v_n=2\cos(n\al+\theta)$ is the appearance of the nonlinear quadratic term $n^2\om$. The conjecture states that the associated Schr{\"o}dinger operator $H$ defined by \eqref{eq:Hdefn} is Anderson localized for arbitrarily small $\lam>0$ everywhere in the spectrum. Partial results in this vein are due to Bourgain \cite{B} and Bourgain-Goldstein-Schlag \cite{BGS}. Note that the conjecture says in particular that the skew-shift dynamics is appreciably more random-like than the circle rotation where $v_n=2\cos(n\al+\theta)$. (Recall that the latter is only localized for $\lam>1$.) The observation that the skew-shift is more quasi-random than the shift has been made in another context by Rudnick-Sarnak-Zaharescu \cite{RSZ} and others \cite{DR,MY,RS} (concerning the spacing distribution) and also recently in \cite{ALY} (concerning eigenvalues of large Hermitian matrices). 

A crucial ingredient for localization on which we will focus is the positivity of the Lyapunov exponent of the associated cocycle, which is defined as follows. From now on, $v_n=v_n(x,y)$ is given by \eqref{eq:vndefn} with $\omega$ irrational.  The second-order difference equation \eqref{eq:schr} can be solved by using transfer matrices:
\beq\label{eq:MAdefn}
\begin{aligned}
\tvector{\psi_{n+1}}{\psi_n}=&M_n(x,y;\lam,E) \tvector{\psi_{1}}{\psi_{0}},\\
\textnormal{where}\quad M_n(x,y;\lam,E):=&\prod_{j=n}^1 A_j(x,y;\lam,E),\\
\textnormal{and}\quad  A_j(x,y;\lam,E):=&\ttmatrix{E-\lam v_j}{-1}{1}{0}.
\end{aligned}
\eeq
The Lyapunov exponent is defined as
$$
L(\lam,E):=\lim_{n\to\infty}\frac{1}{n}\int_{\mathbb T^2} \log \| M_n(x,y;\lam,E)\| \d x \d y,
$$
where the limit exists by subadditivity (Fekete's lemma). We remark that the Furstenberg-Kesten theorem \cite{Via} implies that 
$$
\frac{1}{n}\log \| M_n(x,y;\lam,E)\| \stackrel{n\to\infty}{\longrightarrow} L(\lam,E)
$$
for Lebesgue-almost every initial condition of the skew-shift $(x,y)\in \mathbb T^2$, as long as $\om$ is irrational. This leads us to the following relaxed version of the conjecture from above.

\be{conj}\label{conj}
Let $\om$ be irrational. For every $\lam>0$ and every $E\in \R$, it holds that
$$
L(\lam,E)>0.
$$
\e{conj}

Some limited progress on Conjecture \ref{conj} was made in \cite{B,BGS,HLS1,HLS2,K1,K2}.  We collect some remarks concerning this conjecture.

\be{rmk}
\be{enumerate}[label=(\roman*)]
\item The lower bound on the Lyapunov exponent is nontrivial only when $E$ lies in the spectrum of the Schr{\"o}dinger operator $H$ defined by
\eqref{eq:Hdefn}. A concurrent conjecture, to which we will return later, says that the skew-shift model has no gaps in the spectrum.
\item In the case of i.i.d.\ random $v=(v_n)_{n\in\Z}$ it is known that the Lyapunov exponent is positive by Furstenberg's theorem \cite{Fur}.
\item Herman's subharmonicity trick \cite{H} implies $L(\lam,E)\geq \log \lam$ and thus establishes a lower bound on the Lyapunov exponent if and only if $\lam>1$. The interesting regime for Conjecture \ref{conj} is therefore $\lam\in (0,1]$.
\e{enumerate}
\e{rmk}

\section{A finite-size criterion}
Our paper is a follow-up to an approach on Conjecture \ref{conj} which was recently initiated in \cite{HLS1}. That paper was based on the methods developed in \cite{BGS}, specifically large deviation estimates for Lyapunov exponents and an inductive multi-scale machine based on the Avalanche Principle, and \cite{HLS1} rendered these methods effective, i.e., explicit constants were obtained for every step of the argument. As a result, one obtains finite-size criteria for the validity of Conjecture \ref{conj}. Namely, deriving the conjecture for a fixed choice of parameters $\lam,E$ and $\om=\frac{\sqrt{5}-1}{2}$ (the golden mean) is reduced to verifying $3$ numerical conditions on the Lyapunov exponent at a fixed initial scale, called $N_0$ below. If true, these conditions can be fed into the effective inductive machine from \cite{HLS1} to obtain $L(\lam,E)>0$.

For definiteness, we focus on the following finite-size criterion obtained in \cite{HLS1} (see Theorem 1.4 there). We define the Lyapunov exponent at scale $n\geq 1$ by
\beq\label{eq:Lndefn}
L_n(\lam,E):=\frac{1}{n}\int_{\mathbb T^2} \log \| M_n(x,y;\lam,E)\| \d x \d y
\eeq
and its non-averaged analog by
$$
u_n(x,y;\lam,E):=\frac{1}{n}\log \| M_n(x,y;\lam,E)\|.
$$
We also define the bad set $\curly{B}_n$, where $u_n$ deviates by more than $10\%$ from its average:
\beq\label{eq:badsetdefn}
\curly{B}_n(\lam,E):=\setof{(x,y)\in \mathbb T^2}{\l|u_n(x,y;\lam,E)-L_n(\lam,E)\r|>\frac{L_n(\lam,E)}{10}}.
\eeq

\be{thm}[\cite{HLS1}]
\label{thm:hls}
Let $\om=\frac{\sqrt{5}-1}{2}$ and let $\lam\in [1/2,1]$. Let $N_0=30,000$. Assume that for some energy $E\in [-2-2\lam,2+2\lam]$ the following hold:
\be{enumerate}[label=(\roman*)]
\item $L_{N_0}(\lam,E)\geq 2\times 10^{-3}$,
\item $\frac{L_{N_0}(\lam,E)-L_{2N_0}(\lam,E)}{L_{N_0}(\lam,E)}\leq \frac{1}{8}$,
\item $\max(|\curly{B}_{N_0}(\lam,E)|,|\curly{B}_{2N_0}(\lam,E)|)\leq N_0^{-165}$.
\e{enumerate}
Then:
$$
L(\lam,E)\geq \frac{1}{2} L_{N_0}(\lam,E)\geq 10^{-3}.
$$
\e{thm}

We see that Theorem \ref{thm:hls} reduces the proof of Conjecture \ref{conj} for specific parameters $\lam,\om,E$ to a numerical calculation at the initial scales $N_0=30,000$ and $2N_0$. We remark that the methods in \cite{HLS1} are flexible and can be used to obtain variant finite-size criteria; some further examples are stated in \cite{HLS1}. We work with Theorem \ref{thm:hls} because the initial scale $N_0=30,000$, while large, is amenable to numerical verification which is our goal here. 

Our \emph{main contribution} is to conduct a detailed numerical study of conditions (i)-(iii). The results are obtained by running MATLAB code on Harvard's research computing cluster Odyssey. We mention that numerics for the skew-shift are a delicate matter because of the highly oscillatory matrix elements of $A_j$ defined in \eqref{eq:MAdefn}, particularly the quadratic $n^2\omega$ term in \eqref{eq:vndefn} combined with the irrational nature of $\om$. These difficulties were partially overcome in an earlier numerical study \cite{EBC} which went up to the scale $N=200$. 

The remainder of the paper is organized as follows. In Section \ref{sect:1}, we conduct some preliminary investigations into the spectrum of the Schr{\"o}dinger operator by studying its finite-size approximations and employing some standard mathematical bounds controlling the error of these approximations. In Section \ref{sect:2}, we verify conditions (i) and (ii) of Theorem \ref{thm:hls} numerically for specific parameter choices---see Table \ref{table:lyap} below. In Section \ref{sect:3}, we investigate condition (iii) of Theorem \ref{thm:hls}. Given the large power $165$, condition (iii) cannot reasonably be verified numerically. Nonetheless, we include some graphs that hopefully shed some light on this condition. 

We recall Remark 1.2 in \cite{HLS1} about condition (iii): Condition (iii) is a large-deviation estimate at the initial scales $N_0$ and $2N_0$ and hence plays a different role from conditions (i) and (ii). For instance, analogous large deviations estimates are analytically known to hold a priori, independently of the positivity of the Lyapunov exponent, for the Harper model and others \cite{BG,BGS,GS}. At any rate, an analytical proof of condition (iii) is warranted, and the importance of this problem is elevated further in light of our numerical verification of conditions (i) and (ii) in the present work. 

From now on, we fix the parameters $\omega=\frac{\sqrt{5}-1}{2}$ and $\lam=1/2$.

\section{Numerical results} 
\subsection{Preliminaries on the spectrum of $H$}
\label{sect:1}
As mentioned after Conjecture \ref{conj}, the only energies $E$ for which the positivity of the Lyapunov exponent $L(\lam,E)$ is nontrivial are those lying in the spectrum of the infinite-volume Schr{\"o}dinger operator $H(x,y)$ defined in \eqref{eq:Hdefn} with $v_n=v_n(x,y)$ given by \eqref{eq:vndefn}. General facts from ergodic theory and spectral theory imply that $\mathrm{spec}\,H(x,y)\subset [-2-2\lam,2+2\lam]$ and it does is independent of $(x,y)$ if a zero-measure subset of $\mathbb T^2$ is ignored. 

Apart from these general facts, however, not much is known about $\mathrm{spec}\,H(x,y)$. Hence, it is a priori unclear on which energies $E$ we should focus our numerical investigation. Before we explain how we choose which energies to study (given our choice of other parameters $\omega=\frac{\sqrt{5}-1}{2}$ and $\lam=1/2$), we note that this issue is not as concerning as it may seem at first sight, in light of another standing conjecture that the skew-shift Schr{\"o}dinger operator $H$ given by \eqref{eq:Hdefn} has no gaps in the spectrum. This concurrent conjecture is also supported by numerics \cite{EBC}; see \eqref{eq:gap} below for a numerical upper bound on the spectral gap.

\begin{figure}[t!]
\centering
\includegraphics[scale=.33]{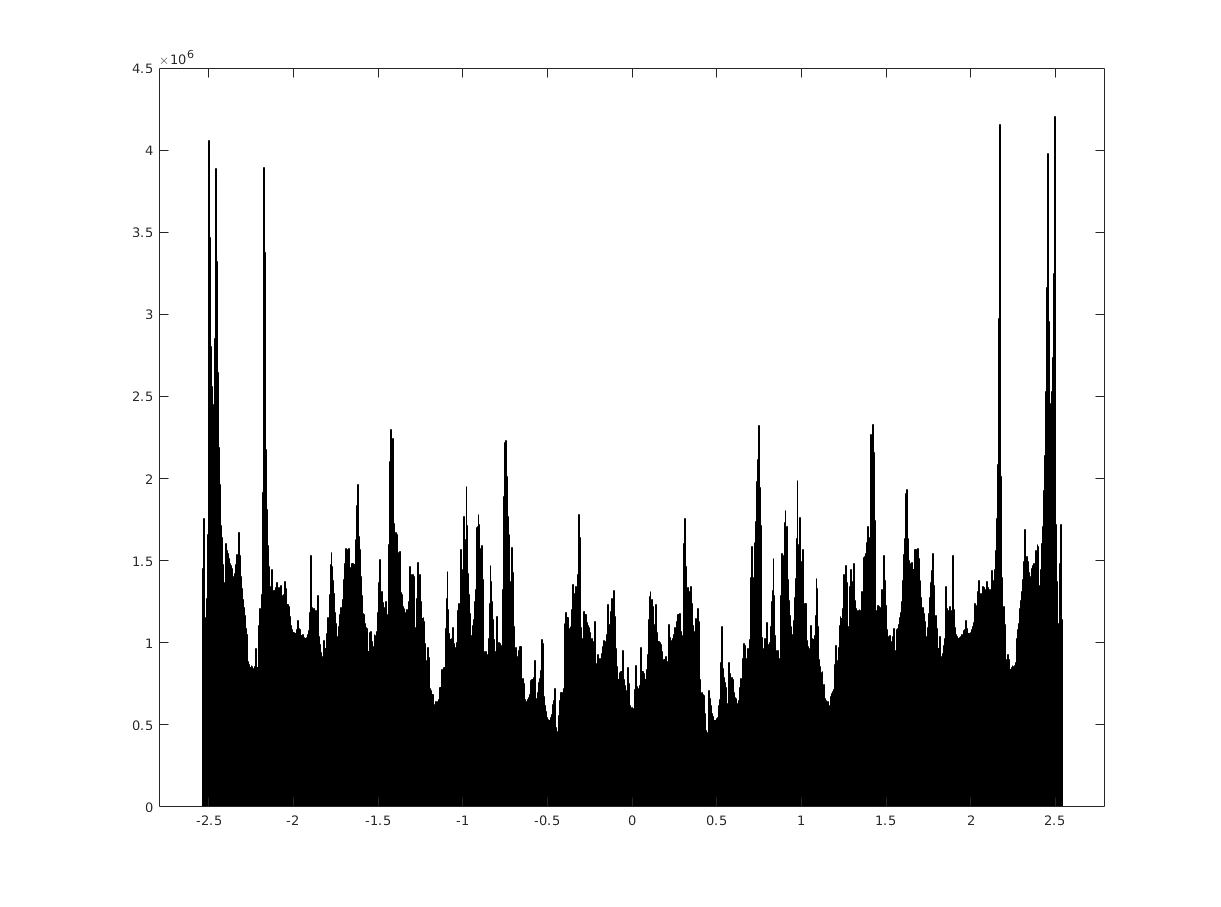}
\caption{\label{fig:batman} A histogram of the eigenvalues of $H_{N_0}(x,y;1/2)$} across a $200\times 200$ regular grid of pairs $(x,y)\in [0,1]^2$. The most common energy when rounding to 8 digits is $E_1=- 2.49512326$; this corresponds to (but is more refined than) the tallest peak on the left in the histogram. A close second is the energy $2.49511562$ which corresponds to the reflected peak on the right.
\end{figure}

We choose to focus our investigations on the energies
\beq\label{eq:Edefn}
E_0=0,\qquad \textnormal{ and } \qquad E_1=-2.49512326.
\eeq
The energy $E_0=0$ is a natural choice by symmetry considerations. Indeed, the spectrum of $H$ is symmetric under reflection at $0$; note  also the reflection symmetry in Figure \ref{fig:batman} for the finite-size analog of this fact.

 The energy $E_1$ is chosen as follows. While $\mathrm{spec}\,H$ is not directly accessible by numerics, we can consider the finite-size approximation to $H$, the $N\times N$ Hamiltonian matrix $H_{N}(x,y;\lam)$ defined by 
$$
H_{N}(x,y;\lam)=\left(\be{array}{cccccc} 
\lam v_1(x,y) &1 & 0 & 0 &  \ldots & 0\\ 
1 & \lam v_2(x,y) & 1 & 0 & \ldots & 0\\
0 & 1 & \lam v_3(x,y) & 1 & \ldots  & 0\\
0 & 0 & 1 & \lam v_4(x,y) & \ddots  & 0\\
\vdots & \vdots & \vdots & \ddots & \ddots  & 1\\
0 & \ldots& \ldots & 0  & 1 & \lam v_{N}(x,y) \\
 \e{array}	\right)
$$
Standard spectral estimates can be used to approximate $\mathrm{spec}\,H$ by $\mathrm{spec}\,H_{N}$, see for instance Corollary 2 in \cite{EBC}. In \cite{EBC}, these estimates are combined with the eigenvalue and eigenvector results for $H_{100}$ to derive the following upper bound on the largest spectral gap in $\mathrm{spec}\, H$:
\beq\label{eq:gap}
\Gam<5.708\times 10^{-4}.
\eeq

We set $N=N_0=30,000$ and diagonalize $H_N(x,y;1/2)$ over a $200\times 200$ regular grid of pairs $(x,y)\in [0,1]^2$. The resulting empirical spectral measure is shown as a histogram in Figure \ref{fig:batman}. We choose $E_1$ as the location of the bin with the tallest peak (using higher precision than the one used to produce the histogram). In other words, $E_1$ is close to an eigenvalue of $H_{N_0}$ for the maximal number of pairs $(x,y)\in [0,1]^2$ and in this sense is the number of maximal likelihood on scale $N_0$ to lie in $\mathrm{spec}\, H$. (We remark that we determine $E_1$ to 16-digit accuracy as $E_1=-2.495123260049612$ and conduct all computations at this accuracy. The approximation \eqref{eq:Edefn} is chosen in the main text for readabiity.)

Moreover, the spectral estimate from Corollary 2 in \cite{EBC} can be used to show that $E_1$ lies inside the spectrum to that level of accuracy, namely,
\beq
\mathrm{dist}(E_1,\mathrm{spec} H)<10^{-8}.
\eeq
This estimate is relevant insofar as we want to check $L(\lam,E)$ for $E\in\mathrm{spec}\, H$ and by the above considerations $E_1$ is our best guess for an element of $\mathrm{spec}\, H$ from spectral information at scale $N_0$.

\subsection{Numerical verification of conditions (i) and (ii)}
\label{sect:2}
We recall that $\omega=\frac{\sqrt{5}-1}{2}$, $\lam=1/2$, and $N_0=30,000$. Our main results are the numerical approximations of the finite-size Lyapunov exponents $L_{N_0}(1/2,E)$ and $L_{2N_0}(1/2,E)$ for $E\in \{E_0,E_1\}$ given in \eqref{eq:Edefn}. The results are summarized in Table \ref{table:lyap}. From these results, we see that
$$
\begin{aligned}
\min_{E\in \{E_0,E_1\}}L_{N_0}(\lam,E)\geq& 8\times 10^{-2}>2\times 10^{-3},\\
\max_{E\in \{E_0,E_1\}}\frac{L_{N_0}(\lam,E)-L_{2N_0}(\lam,E)}{L_{N_0}(\lam,E)}\leq& 2\times 10^{-4} <  \frac{1}{8}.
\end{aligned}
$$
Hence, conditions (i) and (ii) of Theorem \ref{thm:hls} hold numerically for $E_0$ and $E_1$ (in fact, comfortably so).\\

\begin{table}[t]
\centering
\begin{tabular}{|c|c|c|}
\hline
& $E=E_0$ & $E=E_1$\\
\hline
$L_{N_0}(1/2,E)$ & 0.08071& 0.46561 \\
$L_{2N_0}(1/2,E)$ & 0.08070 & 0.46559 \\
$\frac{L_{N_0}(1/2,E)-L_{2N_0}(1/2,E)}{L_{N_0}(1/2,E)}$ &$<2\times 10^{-4}$ & $<5\times 10^{-5}$\\
\hline
\end{tabular}
\vspace{.4cm}
\caption{\label{table:lyap} Finite-size Lyapunov exponents calculated by numerical integration over a regular $2001\times 2001$ grid of $(x,y)\in [0,1]^2$.}
\end{table}

Recall that the Lyapunov exponents $L_{N_0}(1/2,E)$ and $L_{2N_0}(1/2,E)$ are defined as integrals over $\T^2$ in \eqref{eq:Lndefn}. We make some remarks concerning numerical integration errors. We would like to emphasize that there is no viable deterministic bound on the numerical integration error because the integrand $\log \|M_{N_0}(\lam,E;x,y)\|$ can have a large derivative in $x$ and especially $y$. Instead, we independently verified the numbers in Table \ref{table:lyap} with a Monte-Carlo integration using $P=2001^2$ pseudorandom integration points, e.g., we still obtain $L_{N_0}(1/2,E)\approx 0.08071$ in that case. For Monte-Carlo integration the integration error for $L_{N_0}(1/2,E)$ with $E\in \{E_0,E_1\}$ can be estimated as follows. 

The random numerical integration can be studied through the random variable
$$
Q_N(P)=\frac{1}{P}\sum_{i=1}^P u_N(\lam,E;x_i,y_i)
$$
where we assume that the $P$ random variables $(x_1,y_1),\ldots,(x_P,y_P)$ are independently drawn from Lebesgue measure on $[0,1]^2$.

\be{lm}\label{lm:MC} Let $N\geq 1$, $P=2001^2$ and $0<\eps<1$. With respect to uniform probability measure, it holds that
$$
\mathbb P(|Q_{N}(P)-L_{N}|>\eps\times 10^{-2})\leq \frac{1}{300\eps^2}.
$$
\e{lm}
 
For example, if we take $\eps=7$ and use the numerical result $Q_{N_0}(P)\geq 0.08\times 10^{-2}$,  then we find that $\mathbb P(L_{N_0}<10^{-2})<0.05\%$. The bound is also informative for $\eps= 0.1$ in which case it implies $L_{N_0}>0.0797$ with probability $>75\%$. These bounds and their analogs for $L_{2N_0}$ indicate the correctness of the numerical data in Table \ref{table:lyap}.

\be{proof}[Proof of Lemma \ref{lm:MC}]
By submultiplicativity of the norm, $\|AB\|\leq \|A\|\|B\|$, we have
$$
\mathrm{Var}\l(\frac{1}{N}\log \|M_{N}(\lam,E;x,y)\|\r)
\leq \max_j \log \l\|\ttmatrix{E-\lam v_j}{-1}{1}{0}\r\|.
$$
Since
$$
\l\|\ttmatrix{a}{-1}{1}{0}\r\|=\frac{\sqrt{2+a^2+|a|\sqrt{4+a^2}}}{\sqrt{2}},
$$
and $|v_j|\leq 2$, it follows that
$$
\mathrm{Var}\l(\frac{1}{N}\log \|M_{N}(\lam,E;x,y)\|\r)
\leq \log\frac{\sqrt{2+(|E_1|+1)^2+(|E_1|+1)\sqrt{4+(|E_1|+1)^2}}}{\sqrt{2}}.
$$
Using $|E_1|+1\leq 3.5$, the right-hand side is bounded by $4/3$ and so, by independence,
$$
\mathrm{Var}(Q_N(P)-L_N)=\frac{\mathrm{Var}(\frac{1}{N}\log \|M_{N}(\lam,E;x,y)\|)}{P}
< \frac{4/3}{P}<\frac{1}{3}\times 10^{-6}.
$$
The lemma now follows from applying Chebyshev's inequality with deviation $\eps\times 10^{-2}$.
\e{proof}

\subsection{Numerical investigation of condition (iii)}
\label{sect:3}
As mentioned in the discussion after Theorem \ref{thm:hls}, condition (iii) is a large-deviation estimate at the initial scale and plays a different role from conditions (i) and (ii). 

\begin{figure}[t!]
\begin{center}
\begin{minipage}{.5\textwidth}
\includegraphics[scale=.5]{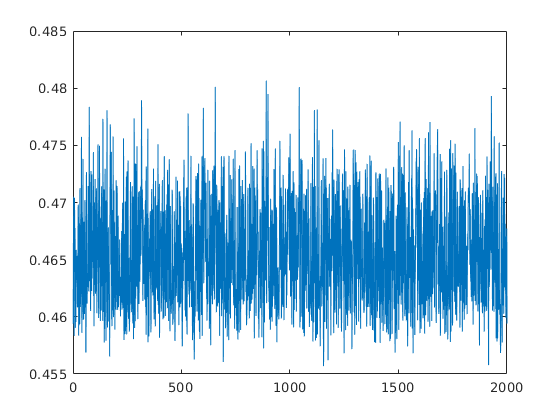}
\end{minipage}\begin{minipage}{.5\textwidth}
\centering
\includegraphics[scale=.5]{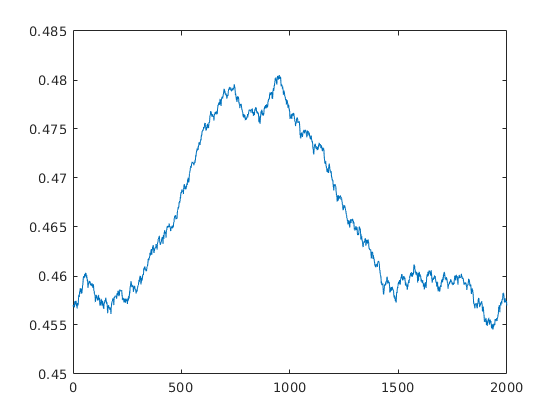}
\end{minipage}
\caption{\label{fig:normal} Example of normal slices of constant $x$- and $y$-values for $N=N_0=30,000$. The energy is $E_0=0$. The left graphs shows the function $y\mapsto L_N(1/2,0;1/2,y)$ and the right graph shows $x\mapsto L_N(1/2,0;x,1/2)$}
\end{center}
\end{figure}

\begin{figure}[t!]
\begin{center}
\begin{minipage}{.5\textwidth}
\includegraphics[scale=.5]{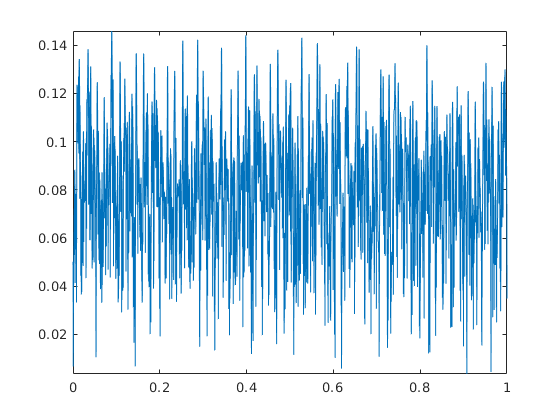}
\end{minipage}\begin{minipage}{.5\textwidth}
\centering
\includegraphics[scale=.5]{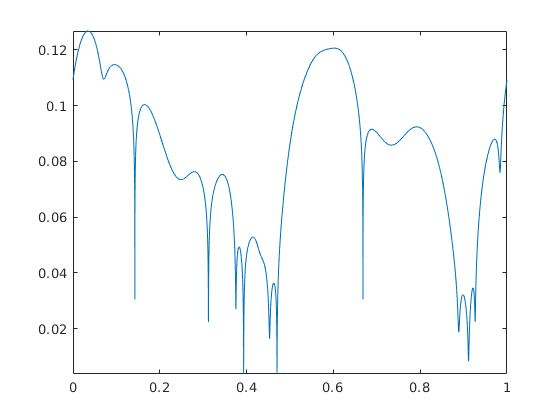}
\end{minipage}
\caption{\label{fig:bad} Example of bad slices for $N=100$. The energy $E_*=0.03688972$ is chosen as an eigenvalue of the finite-size approximation $H_{N}(0.5,0.5)$. The left graphs shows the function $y\mapsto L_N(1/2,E_*;x_0,y)$ and the right graph shows $x\mapsto L_N(1/2,E_*;x,y_0)$ with $x_0=y_0=0.46927639$}
\end{center}
\end{figure}

In this section, we present two different kinds of graphs of the Lyapunov exponent $L_N(1/2,E;x,y)$ along slices of $\T^2$ of constant $x$- and $y$-values for comparison purposes.
\begin{itemize}
\item Figure \ref{fig:normal} shows $L_N(1/2,0;x,y)$ along normal $x$- and $y$-slices. In this case, there is no visible bad set $\curly{B}_N(1/2,0)$.
\item Figure \ref{fig:bad} shows $L_N(1/2,E_*;x,y)$ along slices of constant $x$- and $y$-values which are designed to contain a point in the bad set $\curly{B}_N(1/2,E_*)$ where the Lyapunov exponent dips close to zero. We find such ``bad slices'' by taking the energy $E_*$ to be an eigenvalue of the corresponding finite-size operator $H_N(0.5,0.5)$ and looking for a pair of slices that intersect at a nearby point along which the Lyapunov exponent is minimal. 
\end{itemize}

From both figures, we see that the $x$- and $y$-dependence of $L_N(1/2,E;x,y)$ are starkly different. The fact that the $y$-dependence is more oscillatory is expected in view of the expression $ny+x$ in \eqref{eq:MAdefn}. The main difference between Figures \ref{fig:normal} and \ref{fig:bad}, is that in Figure \ref{fig:bad} the Lyapunov exponent gets close to $0$ at a few special points along both slices. These special points contribute to the bad set $\curly{B}_N$ defined in \eqref{eq:badsetdefn}. 

It turns out that the method we use to construct these ``bad slices'' by hand (described in the second bullet point above) only works for relatively small $N$ and this is the reason why we take $N=100$ in Figure \ref{fig:bad}. This fact can be considered good news in view of condition (iii). Going a step further, we might assume that the only possible reason for the occurrence of a bad point $(x,y)\in \curly{B}_N(\lam,E)$ is that $E$ is an eigenvalue of the finite-size operator $H_N(x,y)$. With this assumption, we can estimate the size of the bad set using the data gathered in Section \ref{sect:1}. Indeed, it suffices to count what is the maximum number of times that an energy $E$ occurs as an eigenvalue of $H_N(x,y)$ for $(x,y)$ ranging over the $2001\times 2001$ toroidal grid. The resulting count, when rounding energies to 8 decimal digits, is $31$. This suggests a rough estimate on $|\curly{B}_{N_0}(1/2,0)|$ of $31/(2001)^2<8\times 10^{-5}$, which is still quite far from what is required for condition (iii).

In light of these results, we believe that finding an analytical proof of condition (iii) makes for an important open problem.

\section*{Supplemental material}
The MATLAB code used to perform the computations presented in this paper and the complete numerical output (to 16-digit accuracy) are available through an online repository \cite{supp}.

\section*{Acknowledgments}
The authors thank Wilhelm Schlag for encouragement and advice. The computations in this paper were run on the Odyssey cluster supported by the FAS Division of Science, Research Computing Group at Harvard University.


\begin{thebibliography}{99}

\bibitem{ALY} Adhikari, A., Lemm, M., and Yau, H.-T.  {\em Global eigenvalue distribution of matrices defined by the skew-shift}, 	arXiv:1903.11514

\bibitem{Anderson}
Anderson, P.W. {\em Absence of diffusion in certain random lattices}, Phys.\ Rev.\ 109 (1958)  (5): 1492--1505. 
 
\bibitem{B}  
Bourgain, J. {\em 
On the spectrum of lattice Schr\"odinger operators with deterministic potential},
Dedicated to the memory of Thomas H. Wolff. 
J.\ Anal.\ Math.\ 87 (2002), 37--75. 

\bibitem{BG} Bourgain, J., Goldstein, M. {\em On nonperturbative localization with quasi-periodic potential},  Ann.\ of Math.\ (2) 152 (2000), no.~3, 835--879.

\bibitem{BGS} Bourgain, J.,  Goldstein, M.,  Schlag, W.  {\em Anderson localization for Schr\"odinger operators on $\Z$ with potentials 
 given by the skew-shift},  Comm.\ Math.\ Phys.\ 220 (2001), no.~3, 583--621. 

\bibitem{CKM}
Carmona, R.,  Klein, A., and Martinelli, F. {\em Anderson localization for
Bernoulli and other singular potentials} Comm.\ Math.\ Phys.\ \textbf{108} (1987), no.\ 1, 41-–66

\bibitem{EBC} Bourgain-Chang, E., {\em Spectral Aspects of the Skew-Shift Operator: A Numerical Perspective}, Comm.\ Comp.\ Phys., 15 (2014), no.\ 3, 712--732

\bibitem{Fur}   F\"{u}rstenberg, H. {\em Noncommuting random products}, Trans.\ Amer.\ Math.\ Soc.~108 (1963), 377--428.

\bibitem{GMP} Goldsheid, I.Ya.; Molchanov, S.A.; Pastur, L.A. {\em A random homogeneous Schr{\"o}dinger operator has a pure point spectrum} Funkcional. Anal. I Prilozen., 11, no.\ 1, 1-10

\bibitem{GS}  Goldstein, M.,  Schlag, W. {\em H\"older continuity of the integrated density of states for quasi-periodic Schr\"odinger equations and averages of shifts of subharmonic functions}, Ann.\ of Math.\ (2) 154 (2001), no.~1, 155--203.

\bibitem{HLS1} Han, R., Lemm, M., Schlag, W. {\em Effective multi-scale approach to the Schr\"odinger cocycle over a skew shift base}, Ergod. Theory Dyn. Syst., https://doi.org/10.1017/etds.2019.19

\bibitem{HLS2} Han, R., Lemm, M., Schlag, W.  {\em Weyl sums and the Lyapunov exponent for the skew-shift Schr{\"o}dinger cocycle}, to appear in J. Spectr. Theory, 	arXiv:1807.00233

\bibitem{DR} Heath-Brown, D.\ R. {\em 
Pair correlation for fractional parts of $\alpha n^{2}$}, 
Math.\ Proc.\ Cambridge Philos.\ Soc.~148 (2010), no.~3, 385--407.

\bibitem{H}   Herman, M.-R. {\em Une m\'ethode pour minorer les exposants de Lyapounov et quelques exemples montrant le caract\`ere local d'un th\'eor\`eme d'Arnol'd et de Moser sur le tore de dimension 2},  Comment.\ Math.\ Helv.\ 58 (1983), no.~3, 453--502. 

\bibitem{Jit}
Jitomirskaya, S. {\em Metal-insulator transition for the almost Mathieu operator}, Ann.\ of Math.\
150 (1999), no. 3, 1159--1175

\bibitem{KS}
Kunz, H., Souillard, B. {\em Sur le spectre des op{\'e}rateurs aux diff{\'e}rences finies al{\'e}atoires}, Comm.\ Math.\ Phys.\ 78 (1980/81), no.~2, 201--246.

\bibitem{K1} Kr\"uger, H. {\em Multiscale analysis for Ergodic Schr\"odinger operators and positivity of Lyapunov exponents. } J.\ Anal.\ Math.\ 115 (2011), 343--387.

\bibitem{K2} Kr\"uger, H. {\em On positive Lyapunov exponent for the skew-shift potential. } (2012) preprint.

\bibitem{MY}
Marklof, J., Yesha, N. {\em Pair correlation for quadratic polynomials mod 1}, Compos.\ Math.\ 154 (2018), no.~5, 960--983. 

\bibitem{RS}
Rudnick, Z.,  Sarnak, P.,
{\em The pair correlation function of fractional parts of polynomials}, 
Comm.\ Math.\ Phys.\ 194 (1998), no.~1, 61--70. 

\bibitem{RSZ}   Rudnick, Z.,  Sarnak, P.,  Zaharescu, A. {\em The distribution of spacings between the fractional parts of $n^{2}\alpha$},  Invent.\ Math.\ 145 (2001), no.~1, 37--57.

\bibitem{supp}
Kielstra, P.M., Lemm, M. {\em Data for ``On The Finite-Size Lyapunov Exponent For The Schrodinger Operator With Skew-Shift Potential''}, https://doi.org/10.5281/zenodo.2638904

\bibitem{Via}  Viana, M. {\em Lectures on Lyapunov exponents}, Cambridge Studies in Advanced Mathematics, 145. Cambridge University Press, Cambridge, 2014.

\end{thebibliography}
\end{document}